# Accurate large-scale simulations of siliceous zeolites by neural network potentials

Andreas Erlebach[1], Petr Nachtigall[1], & Lukáš Grajciar[1]*

[1]*Department of Physical and Macromolecular Chemistry, Charles University, Hlavova 8, 128 43 Praha 2*

The computational discovery and design of zeolites is a crucial part of the chemical industry. Finding highly accurate while computational feasible protocol for identification of hypothetical siliceous frameworks that could be targeted experimentally is a great challenge. To tackle this challenge, we trained neural network potentials (NNP) with the SchNet architecture on a structurally diverse database of density functional theory (DFT) data. This database was iteratively extended by active learning to cover not only low-energy equilibrium configurations but also high-energy transition states. We demonstrate that the resulting reactive NNPs retain DFT accuracy for thermodynamic stabilities, vibrational properties, as well as reactive and non-reactive phase transformations. As a showcase, we screened an existing zeolite database and revealed more than 20k additional hypothetical frameworks in the thermodynamically accessible range of zeolite synthesis. Hence, our NNPs are expected to be essential for future high-throughput studies on the structure and reactivity of siliceous zeolites.

*E-mail: lukas.grajciar@natur.cuni.cz

# 1. Introduction

Zeolites are of central importance for numerous industrial applications ranging from catalysis through adsorption to ion exchange,[1] owing to their highly diverse structures and properties. Theoretically, there are more than two million[2–5] possible zeolite frameworks but only 240 zeolite frameworks listed in the IZA database[6,7] have been prepared so far, a discrepancy known as the zeolite conundrum.[5] Therefore, ongoing research focuses on sophisticated synthesis routes, like the ADOR protocol,[8] allowing the preparation of novel, unfeasible zeolites that are not accessible by standard solvothermal procedures.[9–11] Another way to prepare new feasible or unfeasible zeolites is the polymorphous inter-zeolite transformation under elevated temperature or pressure.[12–16] Finding reliable and computational feasible protocol for identification of hypothetical zeolites that could be synthesized experimentally is still a great challenge.

In order to guide the ongoing search for new zeolites, computer simulations proved indispensable yet challenging for the (pre-)screening of structures and properties. Such a screening performed by Deem et al. allowed to narrow down the number of possible zeolite frameworks to thermodynamically accessible ones.[3,4] This Deem database generated by atomistic simulations using analytical force fields contains about 330k hypothetical zeolites. Other computational studies used the IZA and Deem databases to estimate the feasibility of hypothetical zeolites and formulated design rules for their targeted solvothermal synthesis.[17–20] Central quantity determining the feasibility of zeolites is the correlation between the zeolite density and energy, firstly calculated using atomistic simulations[21] and then confirmed by experiments.[22]

Recently, the advent of machine learning in materials science and chemistry enabled the search for more complex correlations of the zeolite structure,[23] stability, and properties.[24] For example, graph similarity analysis of the Deem and IZA databases predicted thousands of possible diffusionless transformations from known to hypothetical zeolite frameworks.[25] Apart from zeolite synthesis, machine learning studies also used the zeolite databases to find structure-property correlations, *e.g.,* for mechanical properties,[26] discovery of new auxetic materials[27] and gas adsorption capacities to enable the targeted zeolite synthesis.[28] However, the critical prerequisite for finding reliable correlations guiding experimental studies is generating accurate structural and energetic data at the atomistic level.

The atomistic simulations indeed provide vital insights into the structure and properties of zeolites.[29] However, realistic modeling of zeolites with *ab initio* quality is frequently hampered by the prohibitive costs of first-principles methods. For example, only a few studies used atomistic simulations investigating the collapse of zeolites under high temperatures or pressures.[30,31] Under high temperatures and low to moderate pressures, zeolites show a two-stage transformation, first to a low-density and subsequently to a high-density amorphous phase.[31–34] Computational studies of such phase transition used either *ab initio* simulations employing simple structure models with few atoms and short timescales[30] or more realistic structure models and longer timescales but with analytical (reactive) force fields.[31] Large-scale simulations with *ab initio* quality are therefore of fundamental importance for discovering new zeolites not only by the screening of databases but also through the understanding of reaction pathways of zeolite phase transformations.

Enabling such simulations at a large-scale requires approximate modeling of the potential energy surface (PES) that retains the accuracy of high-level quantum mechanical calculations. In recent years, numerous machine learning potentials (MLP) have been proposed that accurately interpolate the PES providing the necessary speed-up compared to *ab initio* simulations.[35–39] Among them are neural network potentials (NNP)[35] of different types and architectures, *e.g.*, hierarchical interacting particle NNP (HIP-NN),[40] tensor field networks,[41] and the graph convolutional NNP SchNet.[42,43] The latter is a message-passing type NNP architecture that uses trainable input representations of atomic environments repeatedly refined by convolutional operations in several iterations to model many-body interactions. Tests on benchmark datasets[37,42,43] focusing on molecular systems proved the very good accuracy of SchNet NNPs to model energy and forces. However, little is known about transferability and accuracy of SchNet for materials science related questions,[36] such as diffusion,[44] phase stability[45] and transitions,[46] phonon properties,[47] and especially its robustness for reactive phase transformations of zeolites. So far, a few studies trained MLPs for PES modeling silica using polymorphs, surface models, amorphous and liquid configurations,[48,49] including a recently trained MLP accurately modeling the structure and high pressure phase diagrams of dense silica polymorphs.[50] However, no silica MLP training considered the tremendous structural diversity of zeolites and their reactive phase transformations.

The central aim of this work is the training of reactive SchNet NNPs for accurate and general PES modeling of silica, including the structural diversity of zeolites over a wide density range. Training of an NNP ensemble allows active learning for iterative extension of the reference dataset and refinement of the NNP.[35,51] The final dataset covers the silica configuration space ranging from low-density zeolites to high-pressure polymorphs, including low-energy equilibrium structures and high-energy transition states. This allows interpolation of the PES for accurate and transferable modeling of siliceous zeolites within the most relevant parts of the configuration space and enables the required large-scale simulations with *ab initio* accuracy.

The trained NNPs facilitated the reoptimization of the Deem database with high accuracy providing vital input for future machine learning studies to find correlations between structure, stability, and properties of zeolites. The database reoptimization also revealed more than 20k additional hypothetical zeolites in the thermodynamically accessible range of zeolite synthesis. In addition, rigorous accuracy tests of the NNPs showed good agreement with DFT and experimental results, including not only equilibrium structures and phonon properties but also silica phase transformations under extreme conditions such as glass melting and the thermal collapse of zeolites. The trained NNPs show an accuracy improvement of about one order of magnitude for modeling energy and forces compared to other PES approximations: two state-of-the-art analytical force fields including the non-reactive Sanders-Leslie-Catlow (SLC) potential[52,53] and the reactive silica force field ReaxFF of Fogarty et al., and one tight-binding DFT parameterization GFN0-xTB.[54,55] Consequently, this work provides a computational tool for accurate, reactive modeling of siliceous zeolites for their targeted design and synthesis.

## 2. Results

### 2.1 Database generation and NNP training

Key prerequisite for the training of NNPs is the generation of a diverse dataset covering the variety of atomic structures and densities of zeolites in both low and high-energy parts of the PES to accurately model structure, equilibrium properties, and phase transitions. This was achieved by the computational procedure depicted in Figure 1. Firstly, a small zeolite subset of ten frameworks was selected from the Deem database by farthest point sampling (FPS) along with SOAP as similarity metric (see Section 4.1) to capture the structural diversity with the least number of configurations (see Supplementary Figure 1). Then, unit cell deformations and MD simulations sampled low and high-energy parts of the PES using the selected zeolites, higher density polymorphs, surface models and amorphous (AM) silica. FPS extracted the most relevant structures from every MD trajectory to reduce the number of required DFT single-point calculations.

After training of an initial NNP ensemble, structure optimizations of the Deem and IZA database along with extrapolation detection using a query-by-committee approach enabled the search for additional frameworks required to sufficiently cover the zeolite configuration space. The sampling of previously unseen transition states included MD simulations for the melting of β-cristobalite, equilibration of liquid silica and the zeolite amorphization (ZA) of Linde Type A (LTA) and Sodalite (SOD). These simulations and NNP retraining on the extended DFT dataset were repeated until no extrapolation was detected. Two reference methods, PBE+D3 and SCAN+D3, were used to train NNPs (see Supplementary Table 1) on the structural database (see Supplementary Figure 2). The corresponding NNPs (and their ensembles) are termed as NNPpbe (eNNPpbe) and NNPscan (eNNPscan), respectively.

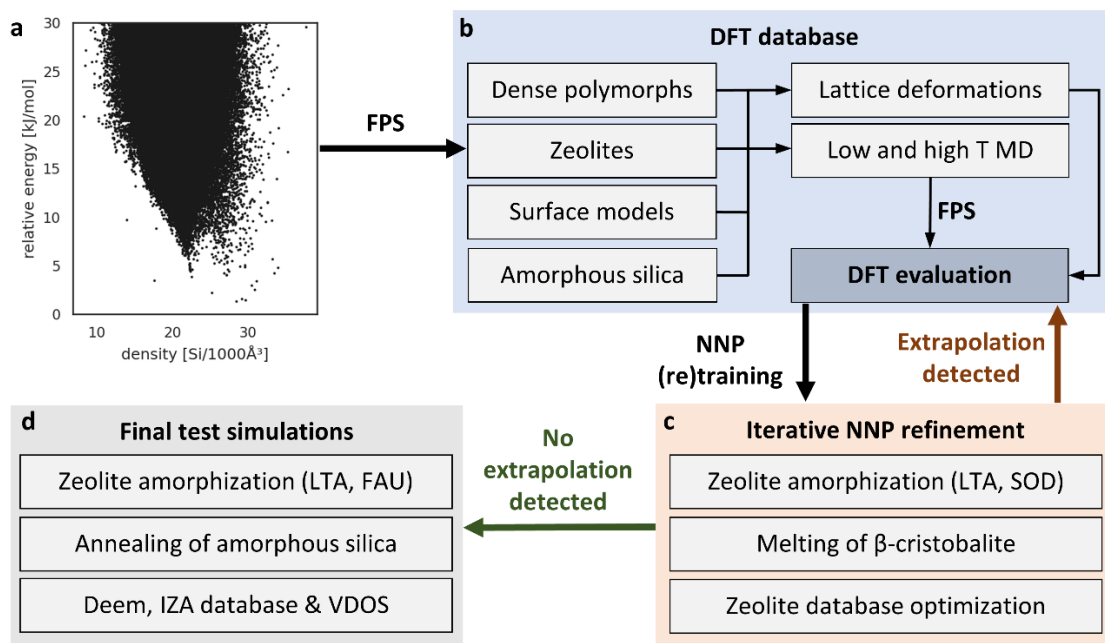

**Figure 1 Computational workflow.** Starting from the Deem database (**a**) containing more than 330k hypothetical zeolites, a structurally diverse subset of zeolites was first selected by farthest point sampling (FPS). Unit cell deformations and MD simulations were applied to the selected zeolites, dense silica polymorphs, surface models and amorphous silica structures (**b**), followed by DFT single-point calculations on an FPS sparsified set of configurations. The NNPs trained on the initial database were iteratively refined by active learning using structure optimizations and MD simulations at high temperatures and pressures (**c**). The final NNP level calculations allowed the reoptimization of the Deem and IZA database, prediction of structure and vibrational properties, and simulations of highly reactive phase transformations such as glass melting and zeolite amorphization, including zeolites (FAU) not present in the training set (**d**).

Finally, NNP level simulations were performed for: (i) the reoptimization of the Deem and IZA database, (ii) glass melting and ZA using Faujasite (FAU) not included in the training set to evaluate the NNP robustness for reactive phase transformations, and (iii) prediction of equilibrium structures and vibrational properties to compare the NNP results with their reference method and experimental data. Further details on the NNPs training and dataset details are provided in Section 4.

When confronting NNPs results with available experimental results, it must be stressed that NNPs cannot outperform the reference method used for generation of the training data. However, accurate NNPs help us understand the accuracy of the reference level of theory (DFT in this work) with respect to experiment providing that the NNPs retain DFT accuracy. Highly accurate NNPs can be used for the simulations of experimental observables using more realistic models and longer simulation times than possibly allowed with more demanding DFT calculations. Thus, the accuracy of the employed DFT methods and the NNPs is demonstrated first (Section 2.2) while the performance with respect to experimental data is described in following subsections (2.3-2.5).

## 2.2 Accuracy compared to other methods

Before evaluating NNP accuracy, we first benchmark the accuracy of the dispersion corrected PBE and SCAN functionals using available experimental data for the structure and energetics of siliceous zeolites (see Supplementary Tables 2 and 3). Two dispersion corrections were considered, namely semiempirical D3 correction by Grimme et al.[56] and more involved many body dispersion (MBD)[57] correction (see Section 4.4 for more details). Previous studies showed that dispersion corrected PBE (both D3 and MBD corrected PBE) agrees best with the experimentally determined structures and phase transition enthalpies of siliceous zeolites as compared to several other GGA exchange-correlation functionals, including also the non-local vdW exchange-correlation (XC) functionals.[58] However, no SCAN benchmark data have been reported so far for siliceous zeolites. Therefore, we used a small test to compare the performance between SCAN and PBE. The resulting relative energies are almost the same for both XC functionals and both dispersion corrections and are in very good agreement with observed phase transition enthalpies with mean average deviations (MAD) of about 2-3 kJ mol$^{-1}$, *i.e.*, within the chemical accuracy. However, SCAN shows slightly better performance than PBE for modeling structural features, for example, with a density MAD of 0.2 Si nm$^{-3}$ (SCAN+D3) versus 0.3 Si nm$^{-3}$ (PBE+D3). In addition, Supplementary Figure 3 compares the energies of the reaction pathway of the Stone-Wales (SW) defect formation[59] in a silica bilayer (*in vacuum*) calculated at the PBE+D3, SCAN+D3, and B3LYP level. The B3LYP has been shown[59] to yield the energy barrier for the first reaction step (for silica bilayer supported on Ru(0001) surface) close to experimental activation energies (about 3-4% deviation) and can be approximated as a reference level of theory also for the SW defect formation in a silica bilayer *in vacuum*. In comparison to the B3LYP reference, the PBE+D3 barriers are slightly underestimated (by up to 0.67 eV - approx. 7% of the highest energy barrier) while SCAN+D3 barriers are slightly overestimated (by up to 0.59 eV - approx. 6% of the highest energy barrier). In light of these results, we conclude that SCAN+D3 provides a consistent, albeit small, performance improvement over dispersion corrected PBE for equilibrium properties and reactive transformations of silica. In addition, earlier benchmarking studies for other systems showed that SCAN+D3 consistently outperforms PBE+D3 not only for equilibrium structures but also for reaction energies and activation barriers.[60] Therefore, SCAN+D3 will be taken as the reference DFT method and NNP trained on the SCAN+D3 data (NNPscan) will be considered as the reference NNP in the following sections, in which we will compare its performance to other PES approximations (analytical force fields, tight-binding DFT, etc.) and experimental data.

The NNP accuracy, together with the accuracy of the commonly used SLC potential, a reactive force field (ReaxFF), and one tight-binding DFT implementation (GFN0-xTB), is evaluated for the set of single-point energy calculations on a test set of structures taken from the NNPscan simulations (Sections 2.3-2.5). This test set contains 1460 configurations including (i) close to equilibrium (EQ) structures randomly chosen from the NNPscan optimized zeolite databases (see Section 2.3), (ii) silica bilayer configurations of the Stone-Wales defect formation (see Section 2.5), and (iii) high-energy structures from the glass melting and ZA simulations (see Section 2.5). These structures were not included in the reference dataset for NNP training. The entire test set can be found in the Zenodo repository (https://doi.org/10.5281/zenodo.5827897).

**Table 1.** RMSE and MAE of energies [meV/atom] and forces [eV/Å] calculated for all test cases and only for equilibrium configurations (EQ) with respect to SCAN+D3 results.

| Level of theory | Energy (EQ) | | Forces (EQ) | | Energy (all) | | Forces (all) | |
|---|---|---|---|---|---|---|---|---|
| | MAE | RMSE | MAE | RMSE | MAE | RMSE | MAE | RMSE |
| sNNPscan | 2.99 | 4.20 | 0.048 | 0.070 | 3.90 | 5.44 | 0.175 | 0.303 |
| eNNPscan | 2.83 | 3.95 | 0.046 | 0.067 | 3.83 | 5.49 | 0.155 | 0.265 |
| SLC | 88.0 | 111 | 2.612 | 3.431 | 207 | 313 | 3.206 | 4.166 |
| ReaxFF | 56.4 | 78.4 | 1.266 | 2.996 | 88.8 | 136 | 2.789 | 8.533 |
| GFN0-xTB | 57.5 | 106 | 0.302 | 0.787 | 127 | 201 | 0.735 | 3.391 |

Table 1 summarizes the MAE and RMSE of energies and forces of all methods with respect to SCAN+D3 results (Supplementary Table 4 shows PBE+D3 results). Figure 2 shows the corresponding energy error distributions. In the case of EQ structures, NNPscan energies are in best agreement with SCAN+D3 calculations with an RMSE of less than 4.2 meV atom$^{-1}$, which are about the same as training errors of 4.7 meV atom$^{-1}$ (see Supplementary Table 1) showing the good generalization capabilities of the NNPs. Using the ensemble average of six NNPs (eNNPscan) provides only minor improvement to the NNP accuracy. The analytical potentials (SLC, ReaxFF) and GFN0-xTB show higher errors by more than one order of magnitude. Such energy errors (~100 meV atom$^{-1}$) translate into uncertainties of zeolite phase stability calculations as described in Section 2.3 (see Figure 3) of about 30 kJ (mol Si)$^{-1}$. The fairly good agreement of SLC with experimental and DFT results (see Supplementary Table 5) applies only to existing siliceous zeolites but not to the rigorous test set including pure silica framework models that cannot be synthesized in their high-silica form (see Section 2.3). Therefore, the trained NNPs provide a sufficiently accurate PES covering much larger configurational space than SLC allowing reliable prediction of zeolite topologies that could be thermodynamically accessible, e.g., via alternative synthesis routes beyond solvothermal methods.

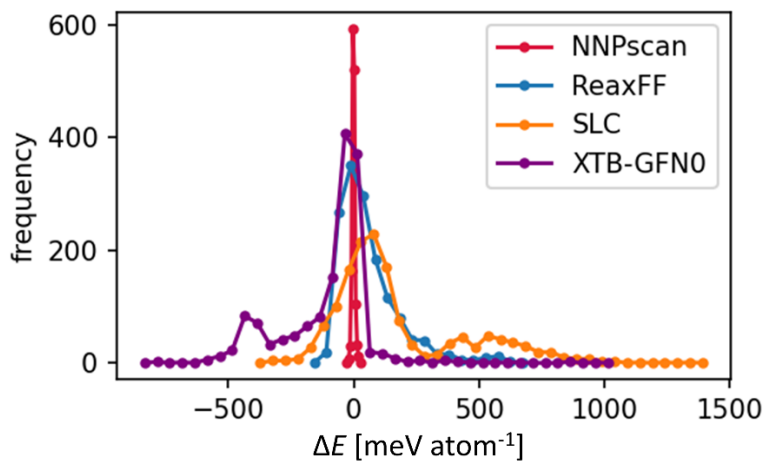

**Figure 2 Error distribution of energies.** Energy errors ΔE are given with respect to SCAN+D3 for analytical force fields (SLC, ReaxFF) and tight binding DFT (GFN0-xTB) as well as NNPscan.

It must be stressed that the trained NNPs approximate energies and forces of the reference level DFT with high accuracy even for high energy structures and transition states (see Table 1). For example, the NNPscan energies deviate about 10-27 meV atom$^{-1}$ from their DFT reference for the glass melting trajectories (see Section 2.5). In addition, even the errors of the extrapolated configurations of the ZA simulations (up to 40 meV atom$^{-1}$) are at least three times lower than the RMSEs of SLC, ReaxFF, and GFN0-xTB. Among the latter, ReaxFF tailored for the elements Si, O, and H provides the lowest energy errors but with an RMSE of 136 meV atom$^{-1}$.[54] Recently, a benchmark study of ReaxFF potentials (parameterized for C, O, H) reported similar energy RMSEs of about 100 meV atom$^{-1}$ for hydrogen combustion reactions.[61] On the other hand, GFN0-xTB allows a more general modeling with a parameterization for 86 elements focusing on equilibrium structures and frequency calculations.[55] Therefore, GFN0-xTB shows a larger energy RMSE for transition state structures (see Figure 2) yet gives higher force accuracy than the silica potentials SLC and ReaxFF. Finally, using an ensemble of six tailor-made silica NNPs provides only a little improvement over single NNP calculations, the latter achieves the best performance in terms of accuracy and computational effort for the reactive modeling of silica.

**2.3 Zeolite databases**

The trained NNPs enable the reoptimization of the Deem and IZA database (available at: https://doi.org/10.5281/zenodo.5827897) to provide highly accurate input for investigations of structure-property relationships of existing and hypothetical zeolites. Figure 3a compares the relative energies and framework densities of the NNPscan optimized databases with the results from the SLC analytical potential, a state-of-the-art analytical potential for silicious zeolites, taken from Ref. 3 (www.hypotheticalzeolites.net, accessed: November 29, 2019). For sake of clarity, only the low-density zeolite analogue RWY[62] ($Ga_2GeS_6$) is not shown in Figure 3 (NNPscan: 61 kJ mol$^{-1}$, 7.86 Si nm$^{-3}$; SLC: 104.2 kJ mol$^{-1}$, 7.62 Si nm$^{-3}$), which does not exist in a high silica form due to a large number of three-membered rings in the structure that would induce high ring tension. Please also note that the Deem database only includes hypothetical zeolites with relative energies up to 30 kJ mol$^{-1}$, which were deemed thermodynamically inaccessible in Ref. 4 and therefore removed from the database.

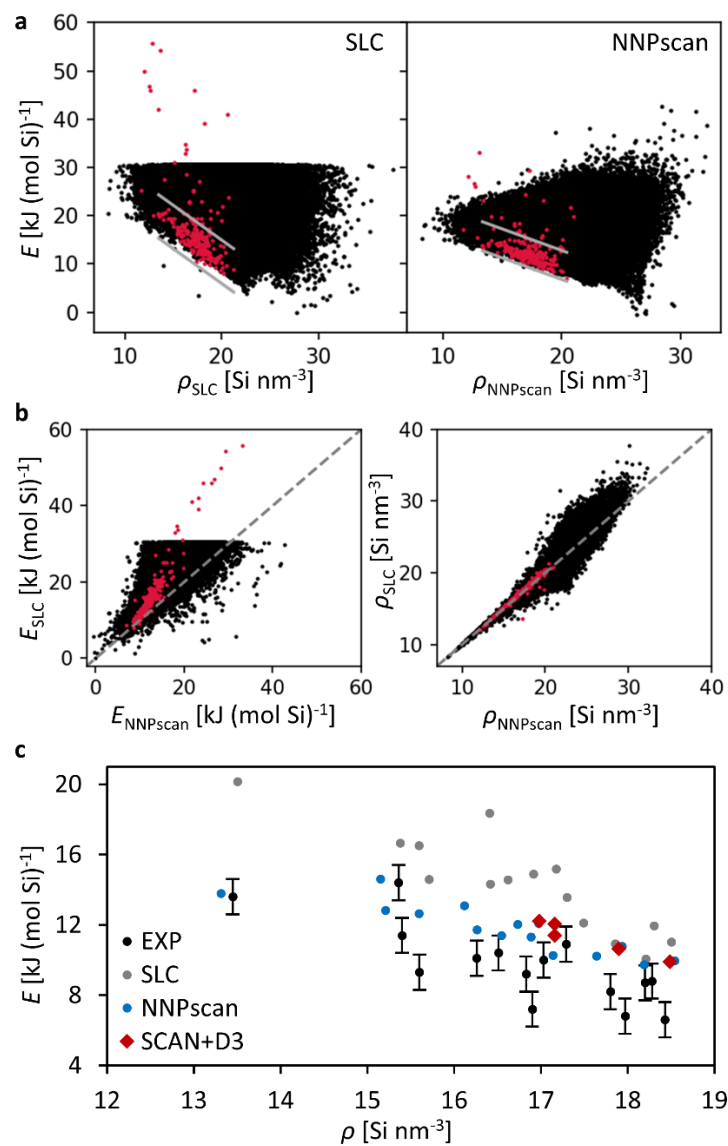

**Figure 3 Calculated energies and densities of siliceous zeolites.** Relative energies $E$ are given with respect to α-quartz as a function of framework density $\rho$ (**a**) of hypothetical (black) and existing (red) zeolite frameworks calculated using SLC and NNPscan. Solid lines in **a** indicate the energy-density range of purely siliceous zeolites (see Supplementary Figure 4). **b** Correlations of energies and densities obtained at the NNPscan and SLC level. **c** Comparison of simulation results with experimental (EXP) values (see Supplementary Table 5). Error bars correspond to an estimated experimental accuracy of ±1 kJ (mol Si)$^{-1}$.[64]

Figure 3b shows the (qualitative) correlation between SLC and NNPscan results - the relative energies (left panel) and densities (right panel) of optimized structures from the Deem and IZA databases are compared. The Pearson correlation coefficients are 0.89 and 0.98 for energies and densities, respectively. However, the SLC results show systematically higher relative energies than NNPscan for zeolites at high energies and densities, probably due to the energetic overestimation of structural features in those zeolites connected with the harmonic three-body bond-bending term of the SLC potential. For example, SLC yielded up to 20 kJ (mol Si)$^{-1}$ higher energies for three-ring containing zeolites such as OBW, OSO, NAB, and JOZ that can only be synthesized if Be is incorporated in the framework (see Supplementary Figure 4).[63] Therefore, the SLC potential probably overestimates the relative energies of hypothetical and existing frameworks that cannot be synthesized as purely siliceous zeolites.

To verify that NNPscan shows improved accuracy compared to SLC, Supplementary Table 5 compares the NNP and SLC results with experimentally available relative enthalpies and densities of 15 siliceous zeolites and five silica polymorphs.[6,7,64] Additionally, DFT optimizations were applied to a subset of five zeolites and five polymorphs. Figure 3c qualitatively compares the dependence of relative energies on the framework density of siliceous zeolites calculated at the SLC, NNPscan, and SCAN+D3 level with experimentally determined phase transition enthalpies and densities. Such energy-density correlations were used in previous studies to find frameworks thermodynamically accessible for solvothermal zeolite synthesis.[3,4] The analytical SLC potential shows relatively good agreement with experimental results in the case of purely siliceous zeolites along with an energy MAE of 4.0 kJ (mol Si)$^{-1}$. However, SLC systematically overestimates the experimentally observed enthalpies (see Figure 3c and Supplementary Table 5), which may relate to the energetic overestimation of structural features in zeolites due to the bond-bending term as described above. In contrast, the trained NNPscan achieved a substantial accuracy improvement with an energy MAE of 2.2 kJ (mol Si)$^{-1}$. Structure optimizations at the SCAN+D3 level of a smaller subset give MAE close to NNPscan, namely, 2.7 kJ (mol Si)$^{-1}$, which is a similar deviation from experiment as reported in previous DFT benchmark studies.[58] The MAEs of atomic densities show a similar trend as that for relative energies, that is, the NNPs provide significantly higher quality than SLC for quantitative structural and energetic predictions of siliceous zeolites with almost same accuracy as dispersion corrected DFT methods (see Section 2.2 and supplementary Tables S2 and S3).

Therefore, reoptimization of the Deem database using NNPscan provides significantly improved input for the computational design and discovery of new zeolite frameworks by analyzing structure, energy, and density correlations for hypothetical and existing frameworks.[3,4,17–20] The solid lines in Figure 3a show the range of relative energies and densities of the 40 zeolite frameworks that have been successfully synthesized in their purely siliceous form (see Supplementary Figure 4).[65] The SLC calculated (relative) energies and densities range between approximately 4-24 kJ (mol Si)$^{-1}$ and 13.5-21.2 Si nm$^{-3}$, respectively. On the other hand, NNPscan optimizations yield a narrower energy range of 6-19 kJ (mol Si)$^{-1}$ but similar densities of 13.3-20.4 Si nm$^{-3}$ (dashed lines in Supplementary Figure 4). Hypothetical zeolites within these energy ranges can be considered as thermodynamically accessible by solvothermal synthesis methods. In the case of SLC, this applies to about 33k frameworks of the Deem database. However, due to the systemically overestimated SLC energies, more than 20k additional hypothetical zeolites (total of about 53k) were obtained from NNPscan calculations that fulfill the stability criterion mentioned above (see Supplementary Figure 5). These results demonstrate the crucial importance of accurate large-scale simulations of equilibrium structures for the discovery of zeolites.

## 2.4 Vibrational properties

In addition to simulations of equilibrium configurations at zero Kelvin, calculations of vibrational properties or free energies at elevated temperatures require accurate modeling of close to equilibrium structures and forces on atoms. To test the reliability of SCAN+D3 and NNPscan for predicting the vibrational density of states (VDOS), the VDOS of α-cristobalite was calculated at both levels. Figure 4a shows both VDOS along with the experimentally observed one[66], and Supplementary Table 6 compares the frequencies of each vibrational mode with the experimental results from IR and Raman spectroscopy.[67–70] Since α-cristobalite was part of the reference database, we performed additional VDOS calculations (NNPscan level) on virtuous silica structures not considered in the NNP training procedure. Three amorphous silica structures were generated using independent simulated annealing MD runs (see Section 4). The obtained VDOS for the glass models are virtually identical (see Supplementary Figure 6), demonstrating sufficient sampling of amorphous structures. Figure 4b shows the average of the three calculated VDOS along with experimental results.[71–73] Please note, the VDOS calculations at the SCAN+D3 level employed the finite-difference approach (FD) using the harmonic approximation while the NNPscan level calculations used MD simulations at 300 K for calculation of the velocity-autocorrelation function (VACF, see Section 4). The latter approach includes anharmonic effects, *i.e.*, the temperature-dependent shift of vibrational frequencies. However, at low temperatures such as 300 K, only minor frequency shifts in the order of 0.1 THz are expected (*e.g.*, as shown before for $Al_2O_3$,[74] $MgSiO_3$[75]) not influencing the comparison of different PES approximations shown in Figure 4.

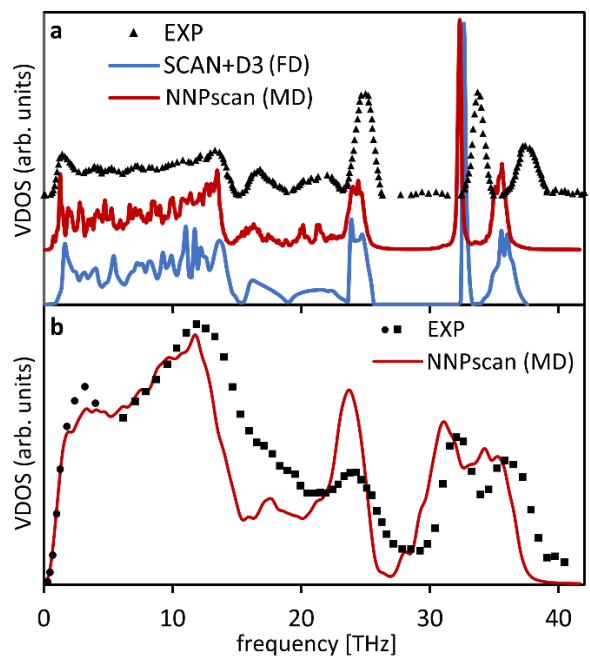

**Figure 4 Vibrational density of states (VDOS).** Comparison of SCAN+D3 and NNPscan calculated VDOS for **a** α-cristobalite and **b** silica glass with experimentally observed VDOS (black triangles: Ref. 66, dots: Ref. 71, squares: Ref. 72, 73). VDOS were calculated using the velocity-autocorrelation from MD simulations (at 300 K) or the finite-difference (FD) approach.

In the case of α-cristobalite, the VDOS calculated at the NNPscan and SCAN+D3 level are almost identical with a frequency MAD of about 0.4 THz for NNPscan compared to the SCAN+D3 reference (see Supplementary Table 6). Both SCAN+D3 and NNPscan show good agreement with experimentally determined frequencies with MADs of 0.3 and 0.5 THz, respectively. The largest deviations of SCAN+D3 (up to 1.6 THz) and NNPscan (up to 2.0 THz) from experiment were obtained for the high-frequency modes $A_2$ (at 34.4 THz) and $B_2$ (at 35.6 THz). We obtained slightly higher frequency errors at the NNPscan level when the FD approach was applied to calculate the harmonic VDOS (see Supplementary Figure 7 and Supplementary Table 6) with an MAD of 0.9 THz from the experiment. These frequency changes are not connected with anharmonic (temperature) effects as described above.[74,75] Most likely, the FD approach is prone to minor force errors of the few single-point calculations required to compute the VDOS. In contrast, the employed MD approach samples the VACF over a trajectory with several thousand microstates, probably facilitating a certain cancellation of the force errors and resulting in a better agreement with the SCAN+D3 FD results.

Since the MD approach proved more accurate for VDOS calculations at the NNPscan level, this procedure was also applied to the VDOS calculations of vitreous silica. Similar to α-cristobalite, the NNPscan calculated VDOS is in good agreement the experimentally observed one. In the case of the high-frequency doublet, the NNPscan calculations yielded a systematic shift by up to 1.5 THz with respect to the experimental VDOS. This shift is similar to the one observed above for the α-cristobalite VDOS at the SCAN+D3 and NNPscan level. Therefore, these systematically underestimated vibrational frequencies are expected to arise from the limitation of the DFT reference method. In summary, the trained NNPs can accurately model equilibrium structures and properties in line with their DFT reference and are in good agreement with available experimental observations.

## 2.5 Phase transitions

Apart from close to equilibrium properties, considering high-energy parts of the PES including transition states is indispensable for simulations of phase transitions and the thermal stability of zeolites potentially leading to the discovery of new zeolites. To showcase the accuracy of the trained NNPs for the description of reactive events the Stone-Wales defect formation[59] in a silica bilayer was chosen as a test case. Figure 5 depicts the reaction path for Stone-Wales defect formation along with DFT and NNP energies (*cf*. Methods section). The bilayer structure is similar to the hypothetical bilayer structure in the reference dataset which consists of four, five, six and ten-membered rings (see Supplementary Figure 2c). However, no transition states from a six to seven-ring topology were included in the training set. Nonetheless, the NNPscan shows good agreement with its DFT reference. NNPscan deviates less than 0.207 eV from the corresponding DFT value, which is about 2% of the highest barrier.

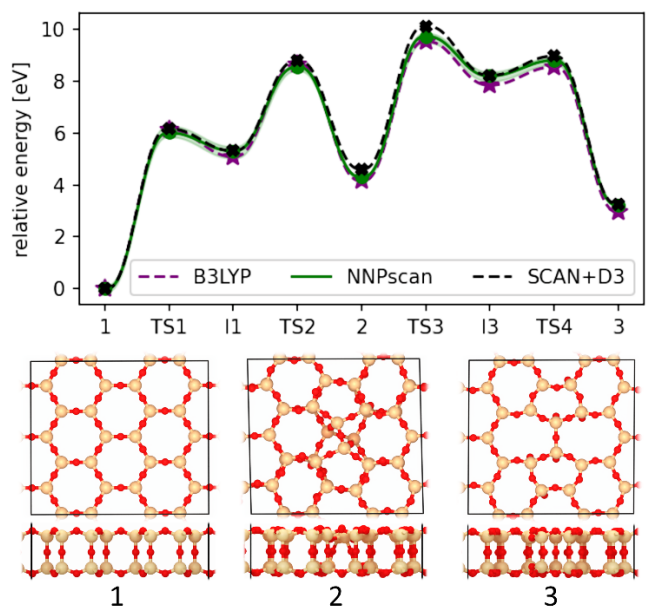

**Figure 5 Energetics of Stone-Wales defect formation.** Energies for transition states (TS) and intermediate (I) structures are calculated at the DFT (B3LYP taken from Ref. 59) and NNPscan level. Shaded areas indicate the energy range calculated with all six NNPs (solid line: ensemble average).

Achieving general modeling of reactive and non-reactive zeolite phase transitions beyond the model reaction path described above requires diverse configurations of the high-energy parts of the PES. Two extreme cases of phase transformations were considered to probe the quality of the PES interpolation between the low-energy EQ and high-energy transition states, *i.e.*, via Si-O bond cleavage (*cf.* Section 4.3): the melting and annealing of amorphous silica and ZA. Figure 6 shows the relative energies with respect to α-quartz for simulations using the NNPs including the melting of β-cristobalite and the amorphization of LTA and FAU zeolites. Note that simulations of β-cristobalite melting and LTA amorphization up to mass density of 2.2 g cm$^{-3}$ (22 Si nm$^{-3}$) were used for training and active learning procedure, however, these simulations were carried out with different potentials, either ReaxFF or the initial NNPs (see Method section). Figure 6 also depicts results of DFT single-point calculations performed for a subset of structures as accuracy checks.

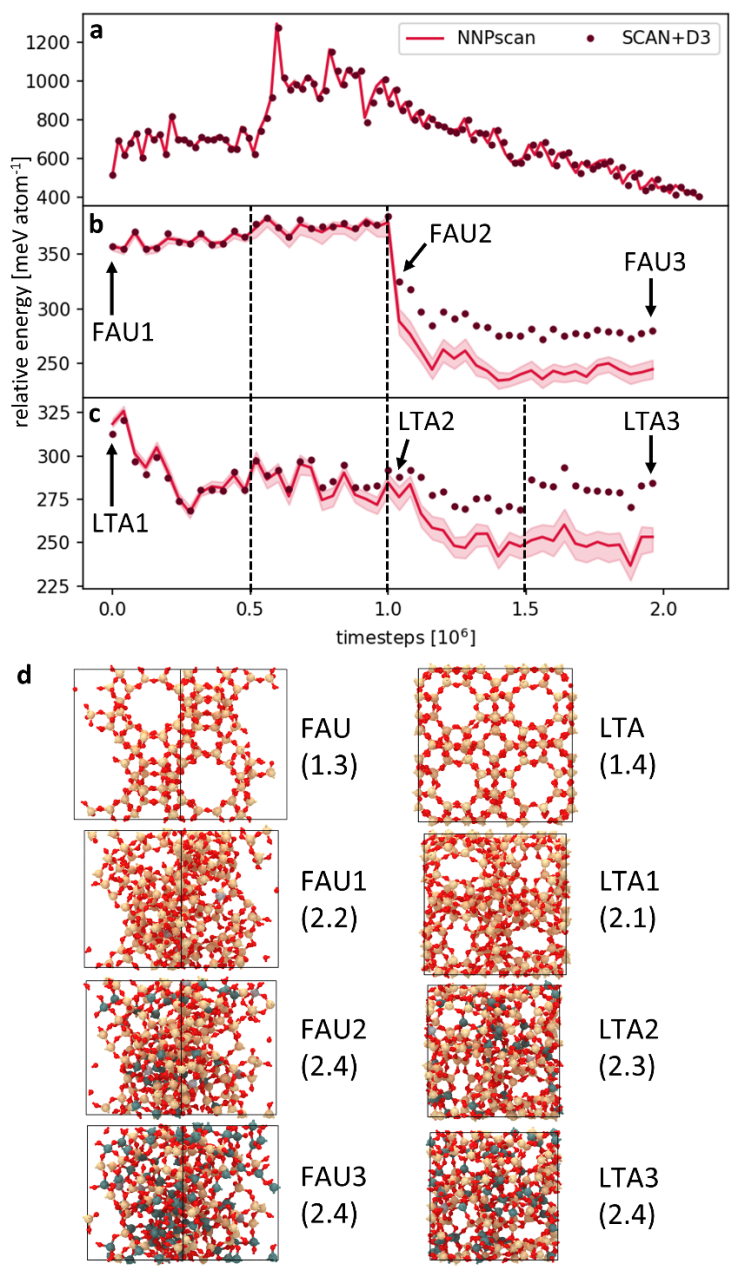

**Figure 6 Reactive silica phase transformations.** Comparison of SCAN+D3 and NNPscan energies for **a** melting and annealing of β-cristobalite as well as amorphization of **b** FAU and **c** LTA by compression. Shaded areas in **a-c** show the energy range of all six NNPs (not visible in **a** due to energy scale) and dashed lines in **b** and **c** indicate compression steps. Structures and mass densities (g cm$^{-3}$, in brackets) are depicted in **d**. Si: yellow, O: red, Si (after bond breaking): green.

During the first $0.5 \cdot 10^6$ timesteps of the β-cristobalite melting simulation at 4800 K, only a few defects were created. The steep increase of the energy at about $0.6 \cdot 10^6$ timesteps corresponds to the phase transition into liquid silica. After $10^6$ timesteps, the temperature was lowered stepwise down to 2500 K with no considerable changes in the structure during the last 100 000 timesteps. Again, NNPscan and SCAN+D3 results show very good agreement. Similar to the results for the Stone-Wales defect formation energies, NNPscan does not deviate from SCAN+D3 by more than 2-3% (~27 meV atom$^{-1}$) .

In contrast to the melting of glass, thermal zeolite amorphization involves not only Si-O bond breaking but also considerable volume changes during the collapse of the framework. To mimic the thermal collapse of LTA and FAU, the structures were equilibrated at 1200 K for 6.5 ns with a stepwise volume reduction every 500 ps such that after 12 equivalent volume steps a mass density of 2.4 g cm$^{-3}$ (24 Si nm$^{-3}$) was reached (*cf*. Method section). The target density exceeds the density range of the configurations in the AM and ZA part of the reference dataset by about 10% to demonstrate the NNP accuracy in extrapolated regions of the PES. Figure 6b and 6c show the energies of the last $2 \cdot 10^6$ timesteps of the trajectories. Figure 6d depicts example structures taken from the MD trajectory. Note that FAU was not included in reference database. In addition, the equilibration time of the last volume step was 1 ns in the case of FAU to ensure full equilibration in the final stage of the framework collapse.

FAU shows no Si-O bond breaking up to the density of amorphous silica (2.2 g cm$^{-3}$). However, the microporous structure considerably changes, mainly due to the collapse of the large cages. Starting from a mass density of 2.4 g cm$^{-3}$, the trajectories of FAU show Si-O bond breakage and reformation events during the last $10^6$ timesteps. The transition states are five-fold coordinated Si leading to cleavage of Si-O bonds and reorientation of SiO$_4$ tetrahedra. The same bond cleavage mechanism was obtained for LTA. However, the first bond breaking was obtained at a density of 2.1 g cm$^{-3}$ as indicated by the energy drop in Figure 6c.

Up to a density of 2.2 g cm$^{-3}$, the deviation of NNPscan from SCAN+D3 is less than 11 meV atom$^{-1}$, which is about 3% of the highest energy at the phase transition. At densities above 2.2 g cm$^{-3}$, larger NNP errors (up to 40 meV atom$^{-1}$) were obtained showing the onset of the extrapolation region in the high-energy part of the PES not covered by the training dataset (see Supplementary Figure 2b). The higher NNP errors are also indicated by the increased spread of the energy predictions of the NNP ensemble enabling future refinement of the NNPs by active learning. Indeed, the NNP energy deviation from the DFT reference considerably increased once the energy spread of the NNP ensemble exceeded ~8 meV atom$^{-1}$ (see Supplementary Figure 8). However, even the extrapolated NNP energies qualitatively agree with the SCAN+D3 results with a Pearson correlation coefficient of 0.99 and 0.66 for FAU and LTA, respectively, showing a fairly systematic energy shift from the reference values (see Supplementary Figure 9). These results demonstrate that SchNet provides reasonable atomic configurations even in extrapolated regions at densities about 10% above the reference data facilitating a robust sampling of the configuration space for further active learning and NNP refinement.

## 3. Discussion

Energy errors of a few meV atom$^{-1}$ and force errors of about 100-300 meV Å$^{-1}$ have been reported previously for state-of-the-art MLP such as moment tensor or Gaussian approximation potentials trained for large-scale simulations of different materials.[38,45–47] Results reported herein show that SchNet NNPs provide the same quality as other MLPs not only for close-to-equilibrium structures of materials but also for high-energy bond-breaking scenarios. In addition, we have demonstrated that the trained SchNet NNPs retain DFT accuracy and provide at least an order of magnitude higher accuracy compared to analytical force fields and tight binding DFT.

Previous trained reactive NNPs for silica[48] used a DFT database containing only two polymorphs (quartz and cristobalite), two surface structures, amorphous, and liquid silica configurations with unit cells comprising less than 144 atoms. The NNPs of Behler and Parrinello show RMSEs of about 200 meV Å$^{-1}$ for forces, *i.e.*, somewhat higher compared to the RMSE of the database test set used in this work (147 meV Å$^{-1}$, see Supplementary Table 1). The DFT database used in this work covers several low- and high-density polymorphs, 2D models, amorphous structures, and the large structural diversity of zeolites using unit cells with up to 400 atoms, including high-energy transition states (see Supplementary Figure 2). Therefore, the NNPs provided in this work aim for a far more general modeling of the silica PES compared to previous studies[48] using only a small number of dense polymorphs, surface models, and amorphous silica structures.

The glass melting simulations clearly demonstrated the good NNP modeling accuracy for bond-breaking events at 4800 K. During the equilibration at such high temperatures, the MD trajectory showed numerous five-fold coordinated transition states of Si in good agreement with DFT results. These MD simulations were performed using silica glass density (2.2 g cm$^{-3}$) covered by the reference dataset containing configurations with densities from about 1.6 g cm$^{-3}$ to 2.2 g cm$^{-3}$ (16-22 Si nm$^{-3}$) for high-energy transition states and densities from 1.0 to about 4.5 g cm$^{-3}$ (10-45 Si nm$^{-3}$) for low-energy EQ structures.

For comparison, the density range of the simulated zeolite collapse was 1.3 to 2.4 g cm$^{-3}$. At densities below 2.2 g cm$^{-3}$, MD simulations showed bond-breaking events in the case of LTA (2.1-2.2 g cm$^{-3}$) and no bond cleavage in FAU. For both zeolites, the NNP energies and forces showed no extrapolation and agreed well with DFT results at densities lower than 2.2 g cm$^{-3}$. Note that FAU was not part of the reference database. Only further compression to artificially high densities up to 10% beyond silica glass density resulted in NNP extrapolation. However, the difference between NNP and DFT energies was even in the extrapolation region at least three times lower (<40 meV atom$^{-1}$) than the RMSEs of the other PES approximations (e.g., 136 meV atom$^{-1}$ for ReaxFF) shown in Table 1. In addition, the MD trajectories contain physically reasonable configurations allowing straightforward extension of the DFT dataset and further NNP refinement. Hence, these ZA simulations demonstrate that the SchNet NNPs are transferable and reasonably data-efficient interpolators of the silica PES as exemplified by qualitatively correct description of zeolite amorphization even slightly beyond the interpolation region.

The employed ZA simulation protocol only mimics the thermal zeolite collapse and does not provide realistic modeling of this phase transition. In fact, there are no reports of the thermal collapse for purely siliceous LTA and FAU. Most experimental studies on such phase transformations used Al-containing zeolites showing that zeolites with Si/Al ratios higher than 4 are thermally very stable due to the higher energetic barrier for breaking Si-O bonds than Al-O bonds.[30,34] Therefore, the ZA simulations required artificial compression to higher densities to obtain a higher degree of amorphization. However, even at lower densities (< 2.2 g cm$^{-3}$), the ZA simulations showed bond-breaking events in LTA without extrapolation and in agreement with DFT results. These results demonstrate that the NNPs also reliably interpolate reactive parts of the PES that are relevant for transformations between different zeolite structures.

In summary, the trained NNPs allow general and accurate modeling of siliceous zeolites with DFT accuracy. This includes modeling of thermodynamic stabilities, equilibrium properties as well as reactive and non-reactive phase transitions of zeolites by interpolation of all *relevant* parts of the PES. Even in the observed cases of extrapolation, the NNPs showed qualitative agreement with DFT results with energy errors far lower than analytical force fields demonstrating the robustness of SchNet NNPs that allows their straightforward refinement and extension by active learning. Thanks to active learning, the NNPs capture the structural diversity of zeolites that is used for re-optimization of the Deem database with high accuracy. The revised database provides vital input for future machine learning studies on structure-stability-property correlations facilitating the computational – *in silico* – design and discovery of zeolites. Finally, NNP extension for modeling zeolites containing heteroatoms such as Al or guest molecules such as water is a promising route towards realistic atomistic modeling of zeolites under synthesis and operating conditions[29] with *ab initio* accuracy.

## 4. Methods

### 4.1 Dataset generation

Generation of the initial DFT datasets used PBE+D3 single-point calculations applied to a diverse set of structures, including silica polymorphs, surface models, hypothetical, and existing zeolites. First, ten hypothetical zeolites were selected from the Deem database by Farthest Point Sampling (FPS)[76,77] to find the most diverse subsample of atomic environments. The FPS employed the similarity distance metric $d(A, B)$ between two zeolites $A, B$ calculated using the average similarity Kernel $\bar{K}(A, B)$ of the smooth overlap of atomic positions (SOAP)[78] descriptor (see Supplementary methods):[79]

$$d(A, B) = \sqrt{2 - 2\bar{K}(A, B)}. \tag{1}$$

Apart from the ten selected zeolites, the FPS detected a hypothetical silica bilayer *in vacuo* (72 atoms, 12 Å vacuum layer), which was also added to the dataset (see Supplementary Figure 2c). Additionally, the dataset included a hypothetical α-quartz (001) surface model (120 atoms, 15 Å vacuum layer) terminated with dangling Si-O bonds. The dataset also contained five existing zeolites (CHA, SOD, IRR, MVY, MTF) and six silica polymorphs (α-quartz, α-cristobalite, tridymite, moganite, coesite, and stishovite) for consideration realistic silica structures.

All selected configurations were optimized at the PBE+D3 level under zero pressure conditions. Next, 210 different unit cell deformations were applied to all optimized structures (see supplementary information). Further sampling of atomic environments close to the optimized configurations used ten ps MD equilibrations (ReaxFF level) at 600 K and 1200 K. The 200 most diverse structures were extracted from every MD trajectory by the FPS described above. The resulting set of structures constitutes the low-energy, close to equilibrium (EQ) part of the silica database (see Supplementary Figure 2).

Sampling of high-energy configurations and transition states used MD simulations (ReaxFF level) for melting and simulating annealing of β-cristobalite (2×2×2 supercell). After scaling its mass density from 2.3 to 2.2 g cm$^{-3}$ (silica glass density) and geometry optimization, the structure was equilibrated for 100 ps at 6000 K. Next, the temperature was reduced to 3000 K in three steps along with an equilibration for 100 ps at each temperature step. The equilibration at 3000 K used additional 100 ps to improve the structural sampling. Again, FPS was applied to the MD trajectories to find the 1000 most diverse configurations. To generate low-energy amorphous structures, ten configurations from the 3000 K MD trajectory were optimized (quenched) at constant volume (PBE+D3 level). The lowest energy structure obtained was optimized under zero pressure conditions. Subsequently, the 210 lattice deformations used above (see supplementary information) were applied to the fully optimized unit cell. Structures generated from the simulated annealing of β-cristobalite are denoted as amorphous silica in the reference database (see Supplementary Figure 2).

In contrast to the melting of silica polymorphs, low-density zeolites show a significant volume contraction during melting, that is, thermal collapse. To mimic this process, eight hypothetical zeolites were equilibrated at 1200 K for 100 ps employing ReaxFF. Then, the unit cell volume was scaled stepwise such that after ten equivalent steps, the mass density of silica glass (2.2 g cm$^{-3}$) was reached. After each contraction step, the zeolites were equilibrated for 100 ps. FPS of the resulting trajectories located 1000 diverse structures for each zeolite that fall into the ZA category of the dataset (see Supplementary Figure 2).

Single-point calculations at the PBE+D3 level were applied to the initial database providing energies and forces for the training of an ensemble of six NNPs allowing their iterative refinement.

### 4.2 NNP refinement

Refinement of the initially trained NNP ensemble requires extrapolation detection for previously unseen configurations. This is achieved by performing simulations using one leading NNP and applying single-point calculations to the trajectory using the remaining five potentials.[35] If the energy and force predictions deviate by more than 10 meV atom$^{-1}$ or 750 meV Å$^{-1}$, respectively, from the NNP ensemble average, additional PBE+D3 single-point calculations were added to the reference database. Simulations and re-training of the NNP ensemble were repeated until no extrapolation was detected during test simulations.

To enhance the structural diversity of the EQ dataset, the Deem (>331k structures, www.hypotheticalzeolites.net, accessed: November 29, 2019) and IZA (235 fully connected frameworks) databases were optimized using constant (zero) pressure conditions. Additionally, β-cristobalite was equilibrated at 4800 K for 1 ns to sample more liquid silica configurations. Extension of the ZA dataset used the same computational protocol for the thermal collapse (up to 2.2 g cm$^{-3}$) of zeolites described above but for the frameworks LTA and SOD, which were not considered in the initial ZA dataset. The resulting PBE+D3 dataset contains about 33k structures with up to 400 atoms per unit cell. Single point calculations at the SCAN+D3 level were also applied to the final database to train the NNPscan potentials. More details on the DFT database are summarized in the supplementary information (Supplementary Figures 1 and 2).

Training of NNPpbe and NNPscan used energy and forces of the databases calculated at the PBE+D3 and SCAN+D3 levels, respectively. The resulting test RMSE are approximately 4.7 meV atom$^{-1}$ for energies and 147 meV Å$^{-1}$ for forces (see Supplementary Table 1). These errors are about an order of magnitude lower compared to other methods approximating the PES of silica (*cf.* Section 2.2).

### 4.3 Test simulations

The final geometry optimization of the Deem and IZA database was performed at the NNPscan level. To test the NNP quality for reactive events, structures of the Stone-Wales defect formation were taken from Ref. 59. The unit cell parameters were optimized at the NNPscan and NNPpbe level keeping the fractional coordinates and the vacuum layer frozen, followed by DFT single-point calculations using the optimized structures. MD simulations (timestep 1 fs, NNPscan level) of the thermal LTA and FAU collapse served as a test case of the final NNPs (FAU was not part of the final reference dataset). These simulations used the same procedure described above, but with 12 compression steps up to a mass density of 2.4 g cm$^{-3}$ and an equilibration time of 500 ps between each step.

Test simulations (NNPscan level) for the annealing of amorphous silica used three different initial structures: β-cristobalite and two vitreous silica structures taken from the ReaxFF simulated annealing described above. Melting of β-cristobalite employed an equilibration for 1 ns at 4800 K. After geometry optimization, the two amorphous structures were equilibrated for 1 ns at 4200 K due to their lower energetic barrier for transition to the liquid state. In all three cases, the temperature was stepwise decreased to 2500 K in 100 K steps and an equilibration time of 25 ps per temperature step. The last structures of the MD trajectories were optimized under zero pressure conditions. The obtained glass configurations were equilibrated for 10 ps at 300 K (NVT ensemble), followed by another 10 ps equilibration using the NVE ensemble. Calculation of the VDOS used the velocity auto-correlation function from the NVE trajectory. In the case of α-cristobalite, the harmonic VDOS was calculated at the NNPscan and SCAN+D3 level using a 3×3×2 supercell and the finite-difference (FD) approach. In addition, calculation of the anharmonic α-cristobalite VDOS at the NNPscan level employed MD simulations at 300 K with the same computational protocol used for the silica glass structures.

### 4.4 Computational details

DFT simulations at the GGA (PBE)[80] and meta-GGA (SCAN)[81] level employed the Vienna Ab initio Simulation Package (VASP, version 5.4.4)[82–85] along with the Projector Augmented-Wave (PAW) method.[86,87] Calculations at constant volume used a plane-wave energy cutoff of 400 eV while the constant pressure optimizations used cutoff of 800 eV. The **k**-point grids had a linear density of at least one **k**-point per 0.1 A$^{-1}$ along the reciprocal lattice vectors. The consideration of long-range dispersion interactions is essential for accurate modeling of zeolites.[88,58,89] However, it has been shown that accurate modelling of dispersion in porous materials can be challenging[90,91] and that the SCAN functional in particular can exhibit non-systematic accuracy for description of dispersion in systems with variable sizes and densities.[92] Therefore, we considered two types of dispersion corrections with both PBE and SCAN functionals, a simple semiempirical one proposed by Grimme *et al*. (D3)[56] (with Becke-Johnson damping)[93] and more involved density-depended many-body dispersion (MBD)[57] correction, and we compared their performance with available experimental data for equilibrium structures and energies of siliceous zeolites (see supplementary Tables S2 and S3). The MBD correction was used with the vdW scaling parameters $β$ of 0.84 for PBE and 1.12 for SCAN as optimized in Ref. 92. We found that both dispersion corrections provide virtually the same quality with respect to experimental data on equilibrium structures and energies, similar to the results of a previous study (considering PBE functional only).[58] Therefore, we opted out for the computationally less demanding Grimme D3 dispersion correction for the dataset generation.

Training of SchNet[42] NNPs employed the Python package SchNetPack[43] and random splits of the reference datasets into training, validation, and test sets at a ratio of 8:1:1 that showed lowest RMSEs for different tested split ratios (see Supplementary Figure 10). Mini-batch gradient descent optimization was applied for training along with a mini-batch size of eight structures and the ADAM optimizer.[94] During NNP training the learning rate lowered stepwise (from $10^{-4}$ to $10^{-6}$) by a factor of 0.5 if the validation loss shows no improvement after 15 epochs. We used the same squared loss function for energy and forces as in Ref. 42 along with a trade-off factor of 0.01, that is, with high weight on force errors. The setup of the NNP hyper-parameters used six interaction blocks, 128-dimensional feature vectors, a cutoff radius 6 Å and a grid of 60 Gaussians for expansion of pairwise distances as input for the filter generating networks. A similar training and hyper-parameter setup provided very good NNP accuracy and training performance in previous works.[42,43]

Calculations with the trained NNPs employed the atomic simulation environment (ASE).[95] Simulations at the ReaxFF[54] level used the large-scale atomic/molecular massively parallel simulator (LAMMPS)[96,97] and in the case of the Sanders-Leslie-Catlow (SLC) potential[52,53] the General Utility Lattice Program (GULP).[98] GFN0-xTB[55] calculations were performed with the xTB program package (version 6.3.3, available at: https://github.com/grimme-lab/xtb).

Unless stated otherwise, all MD simulations used a time step of 0.5 fs and the canonical (NVT) ensemble with the Nosé-Hoover thermostat.[99,100] Calculation of the harmonic VDOS at the SCAN+D3 and NNPscan level employed the finite-difference (frozen-phonon) approach implemented in Phonopy[101] along with displacements of 0.02 Å. The calculation of the VDOS and anharmonic vibrational frequencies from MD trajectories used the Python packages pwtools (available at: https://github.com/elcorto/pwtools) and DynaPhoPY,[102] respectively. Calculations of the SOAP descriptor were performed with the Python package Dscribe.[103]

## 5. Data availability

The Deem and IZA database, the trained NNPs, and the test set used for accuracy evaluation is openly available in a Zenodo repository (https://doi.org/10.5281/zenodo.5827897). The Deem database contains more than 331k hypothetical zeolite frameworks geometrically optimized at the NNPscan level. The NNPscan optimized database of the International Zeolite Association contains 236 exiting, fully connected zeolite frameworks. Both databases are SQLite database files of the Atomic Simulation Environment (ASE) containing the ASE Atoms objects with energies and forces calculated at the NNPscan level and are readable with ASE's I/O module. Additionally, the repository contains plain text data as csv files containing zeolite features such as relative energies and densities for both the Deem and IZA database. The remaining data for the reproduction of results is available upon reasonable request.

## 6. Code availability

Source code files for reproduction of results are available upon reasonable request.

## Acknowledgments

The authors acknowledge Charles University Centre of Advanced Materials (CUCAM) (OP VVV Excellent Research Teams, project number CZ.02.1.01/0.0/0.0/15_003/0000417), the support of Primus Research Program of the Charles University (PRIMUS/20/SCI/004), and of the Czech Science Foundation (20-26767Y). PN acknowledges the Czech Science Foundation (grant No. 19-21534S).

This work was supported by The Ministry of Education, Youth and Sports of the Czech Republic through the e-INFRA CZ (ID:90140).

## Author contributions

LG and PN acquired funding. LG and AE conceptualized and administered the project. AE generated, curated and analyzed computational data. LG helped with data analysis. AE prepared all the article graphics. LG supervised the investigation. AE wrote the original article draft. LG, PN and AE carried out the manuscript review and editing.

## Competing interests

The authors declare no competing financial or non-financial interests.

*Supplementary information*

# Accurate large-scale simulations of siliceous zeolites by neural network potentials

Andreas Erlebach[1], Petr Nachtigall[1], & Lukáš Grajciar[1*]

[1]*Department of Physical and Macromolecular Chemistry, Charles University, Hlavova 8, 128 43 Praha 2*

## 1. Supplementary methods

The selection of diverse zeolite frameworks from the Deem database used a subset of 69079 structures containing less than 80 atoms per unit cell. The selected subset covers the whole energy-density range of the full database (see Supplementary Figure 1a). To find diverse atomic environments in the subset, the applied farthest point sampling (FPS) used the smooth overlap of atomic positions (SOAP) descriptor as similarity metric.

Calculation of the SOAP power spectrum $\mathbf{p}(\chi_i^A)$ for $n$ atomic environments $\chi_i^A$ in structure A used a cutoff $r_{\mathrm{cut}}$ = 6 Å, $n_{\max}$ = 8 radial basis functions (gaussian type orbitals) and spherical harmonics up to degree $l_{\max}$ = 8. A computationally efficient way to compare two different zeolites is the calculations of the average power spectrum over all atoms in the structure:

$$\widehat{\mathbf{p}}(A) = \frac{1}{n} \sum_i^n \mathbf{p}(\chi_i^A). \qquad (1)$$

The average SOAP similarity kernel $K(A, B)$ of two different structures A, B equals the dot product of the average power spectra $K(A, B) = \widehat{\mathbf{p}}(A) \cdot \widehat{\mathbf{p}}(B)$.[7,8] Calculation of the similarity distance metric (eq 1) for the FPS used the normalized similarity kernel $\overline{K}(A, B)$:[8]

$$\overline{K}(A, B) = \frac{K(A, B)}{\sqrt{K(A, A) K(B, B)}}. \qquad (2)$$

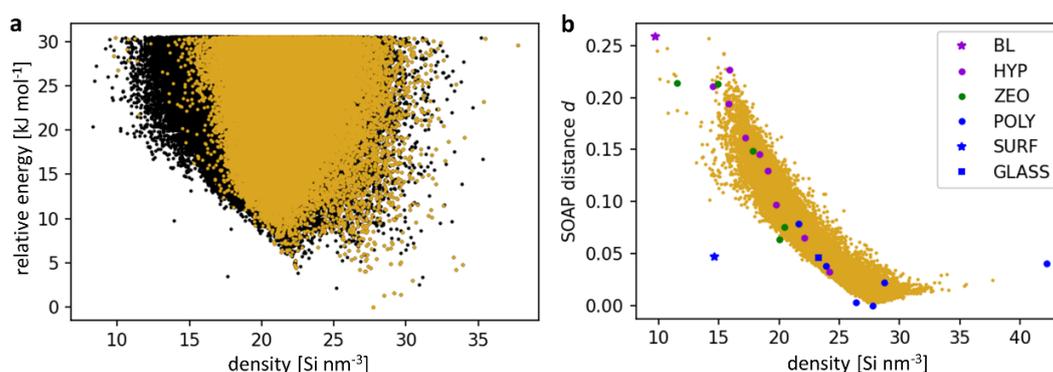

**Supplementary Figure 1 a** Relative energy and **b** SOAP distance *d* (eq 1) with respect to α-quartz as a function of framework density of the Deem database (black dots) and a subset of structures containing less than 80 atoms per unit cell (yellow dots). The equilibrium structures selected for dataset generation are highlighted in **b**: hypothetical (HYP) zeolites from FPS of the subset, existing zeolites (ZEO) and silica polymorphs (POLY), amorphous silica (GLASS), a silica bilayer (BL), and an α-quartz (001) surface model (SURF).

Supplementary Figure 1b shows the SOAP similarity distance (eq 1) as a function of framework density for hypothetical zeolites selected by FPS and existing zeolites, silica polymorphs, silica glass, an α-quartz surface, and a silica bilayer model. The chosen set of structures represents the diversity of densities and atomic environments of silica, ranging from low-density zeolites and surface models to high-density polymorphs such as coesite and stishovite.

To sample diverse configurations close to equilibrium (EQ), the selected structures were first optimized at the DFT level under constant (zero) pressure conditions. Next, DFT single-point calculations were applied to 210 different lattice deformations and 200 configurations extracted by FPS from MD trajectories (ReaxFF level) at 600 and 1200 K for every optimized structure. The deformed unit cells $\mathbf{M}_d$ were obtained using $\mathbf{M}_d = (\mathbf{I} + \boldsymbol{\varepsilon})\mathbf{M}_0$ from the optimized unit cell matrix $\mathbf{M}_0 = (\mathbf{a}, \mathbf{b}, \mathbf{c})$ with the lattice vectors $\mathbf{a}, \mathbf{b}, \mathbf{c}$. The unit cell remains unchanged if the six independent elements $\varepsilon_{ij}$ of the symmetric 3×3 matrix $\boldsymbol{\varepsilon}$ equal zero. There are 70 different permutations for lattice deformations ranging from one single $\varepsilon_{ij} = \pm d$ (rest zero) up to all six $\varepsilon_{ij} = \pm d$. Here, three deformation factors $d = 0.015, 0.03, 0.045$ were considered yielding 210 different lattice deformations.

The NNPs presented here are designed for accurate description of diverse atomic structures and densities of zeolites, covering also high-energy parts of the PES. This was achieved by iterative extension of an initial dataset and (re)training of an NNP ensemble using active learning (see Section 4.1 and 4.2). Supplementary Figure 2 shows the relative energies with respect to α-quartz of the final zeolite database as a function of atomic density and the similarity distance metric $d$ (eq 1) calculated using the smooth overlap of atomic positions (SOAP) descriptor. The database contains low-energy, close to equilibrium (EQ) structures and high-energy structures from phase transition simulations (*cf.* Section 2.4) at high temperatures (amorphous and liquid silica (AM)) and high pressures (zeolite amorphization (ZA)).

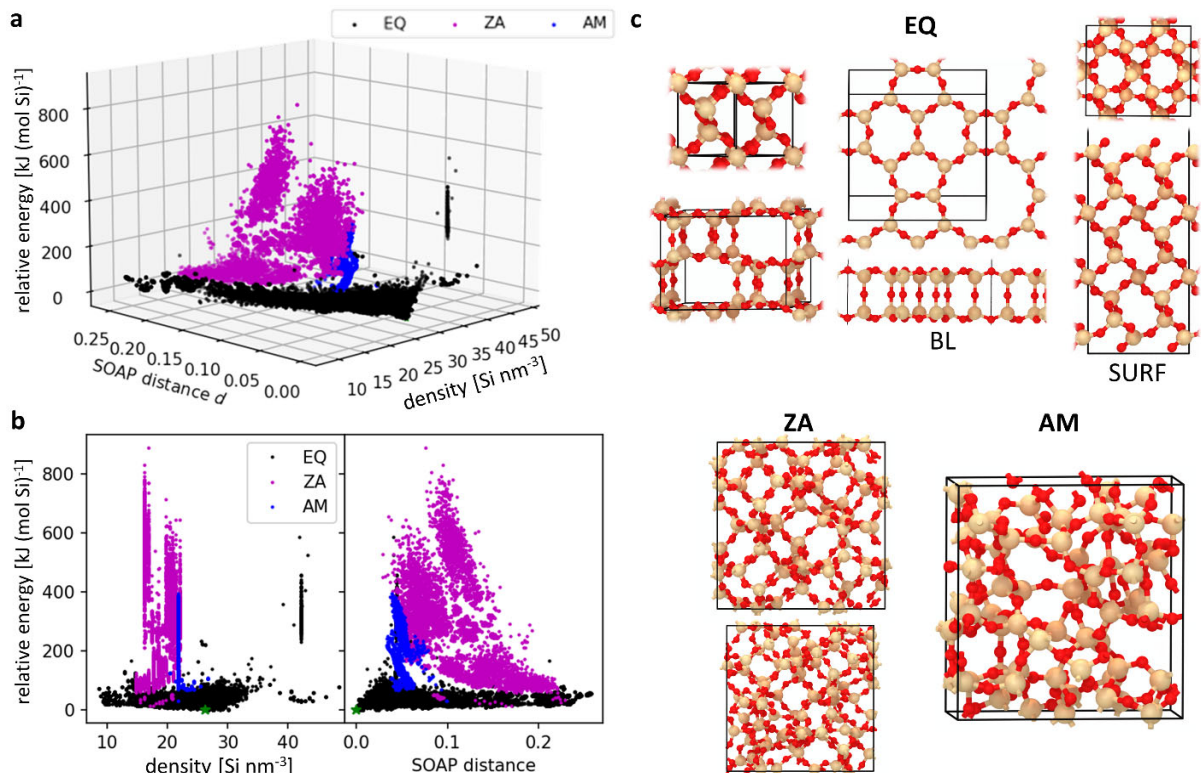

**Supplementary Figure 2 a** Relative energies with respect to α-quartz (green star) of the final SCAN+D3 database as a function of atomic density and SOAP distance $d$ (eq 1), **b** projections of **a**, and **c** example structures close to equilibrium (EQ) including a silica bilayer (BL), and an α-quartz (001) surface model (SURF), amorphous and liquid silica (AM), and configurations from zeolite amorphization (ZA). Si: yellow, O: red.

The EQ part of the database covers atomic structures and densities ranging from low-density zeolites to high-density polymorphs such as coesite and stishovite (Supplementary Figure 2b) to represent the structural diversity of existing and hypothetical silica configurations. It also contains hypothetical surface models, including a silica bilayer (BL) *in vacuo* and an α-quartz (001) surface model (SURF) showing dangling Si-O bonds (Supplementary Figure 2c). Most structures in EQ show tetrahedrally coordinated Si, except stishovite with six-fold coordinated Si. Structures with five-fold coordinated Si are part of AM corresponding to transition states in the liquid state. However, AM includes only densities close to silica glass density (2.2 g/cm$^3$ or 21.8 Si nm$^{-3}$). In contrast, ZA also contains high-energy structures and transition states (five-fold coordinated Si) at lower densities between 16-22 Si nm$^{-3}$ to interpolate the PES for simulations of the thermal collapse of zeolites. Most high-energy configurations show rather large SOAP distances from 0.05 to 0.2, *i.e.*, low structural similarity compared to α-quartz. At lower SOAP distances are mainly higher density structures such as silica polymorphs with relative energies of less than 200 kJ/(mol Si) except for stishovite EQ configurations at densities larger than 40 Si nm$^{-3}$ (*cf.* Supplementary Figure 2b). Therefore, it is expected that the resulting NNPs inaccurately model reactive (high-energy) transformations of high-density silica phases which are beyond the scope of this work. Here, we focus on the accurate, reactive modeling of zeolites including low-density frameworks such as FAU (13.5 Si nm$^{-3}$) as described in Section 2.4.

Supplementary Table 1 summarizes the test set errors of the NNP ensembles trained on the final DFT (PBE+D3 and SCAN+D3) datasets.

**Supplementary Table 1.** Mean absolute (MAE) and root mean square error (RMSE) of the test sets.

| NNPscan | Energy [meV atom$^{-1}$] MAE | RMSE | Forces [meV Å$^{-1}$] MAE | RMSE | NNPpbe | Energy [meV atom$^{-1}$] MAE | RMSE | Forces [meV Å$^{-1}$] MAE | RMSE |
|---|---|---|---|---|---|---|---|---|---|
| 1 | 2.83 | 4.63 | 79.51 | 148.47 | 1 | 2.48 | 4.01 | 69.71 | 128.15 |
| 2 | 2.79 | 4.62 | 79.73 | 151.42 | 2 | 2.70 | 4.84 | 78.03 | 149.36 |
| 3 | 2.84 | 4.75 | 80.27 | 149.52 | 3 | 2.71 | 5.15 | 73.64 | 145.81 |
| 4 | 2.80 | 4.89 | 80.30 | 155.31 | 4 | 2.67 | 5.37 | 74.22 | 146.37 |
| 5 | 2.85 | 4.59 | 85.21 | 156.18 | 5 | 2.58 | 4.86 | 73.17 | 140.77 |
| 6 | 2.73 | 4.30 | 79.21 | 144.96 | 6 | 2.77 | 4.77 | 73.70 | 141.49 |

## 2. Supplementary discussion

**Supplementary Table 2.** Relative energies ΔE [kJ (mol Si)⁻¹], density ρ [Si nm⁻³], average Si-O bond distances d(Si-O) [Å] and mean average deviations (MAD) of lattice parameters ΔL [Å] from experiments calculated at the PBE level with two dispersion corrections (D3, MBD), and NNPpbe level (PBE+D3) for α-quartz (qu), α-cristobalite (cr), tridymite (tri) and purely siliceous zeolites. Experimental values for relative transition enthalpies ΔH and structural parameters were taken from Refs. 3–5.

|     | EXP |   |   | PBE+D3 |   |   |   | PBE+MBD |   |   |   | NNPpbe |   |   |   |
| --- | --- | --- | --- | --- | --- | --- | --- | --- | --- | --- | --- | --- | --- | --- | --- |
|     | ΔH | ρ | d(Si-O) | ΔE | ρ | d(Si-O) | ΔL | ΔE | ρ | d(Si-O) | ΔL | ΔE | ρ | d(Si-O) | ΔL |
| qu  | 0.0 | 26.52 | 1.61 | 0.0 | 25.92 | 1.62 | 0.04 | 0.0 | 25.91 | 1.62 | 0.04 | 0.0 | 25.11 | 1.62 | 0.10 |
| cr  | 2.8 | 23.38 | 1.60 | 4.0 | 22.68 | 1.62 | 0.05 | 3.7 | 22.74 | 1.62 | 0.05 | 4.3 | 24.22 | 1.63 | 0.10 |
| tr  | 3.2 | 22.35 | 1.56 | 6.5 | 21.61 | 1.62 | 0.13 | 4.9 | 21.61 | 1.62 | 0.13 | 7.1 | 21.75 | 1.62 | 0.15 |
| AFI | 7.2 | 16.89 | 1.61 | 12.0 | 16.89 | 1.62 | 0.04 | 10.4 | 16.76 | 1.62 | 0.03 | 11.6 | 16.77 | 1.62 | 0.03 |
| FER | 6.6 | 17.55 | 1.61 | 11.2 | 17.46 | 1.62 | 0.03 | 9.3 | 17.48 | 1.62 | 0.02 | 11.3 | 18.34 | 1.62 | 0.19 |
| IFR | 10.0 | 17.15 | 1.61 | 12.0 | 17.15 | 1.62 | 0.10 | 11.0 | 16.72 | 1.62 | 0.11 | 12.0 | 16.78 | 1.63 | 0.09 |
| MTW | 8.7 | 18.23 | 1.61 | 10.9 | 18.23 | 1.62 | 0.06 | 8.8 | 18.03 | 1.62 | 0.06 | 11.4 | 18.17 | 1.62 | 0.11 |
| **MAD** | - | - | - | 3.0 | 0.30 | 0.02 | 0.06 | 1.6 | 0.40 | 0.02 | 0.06 | 3.2 | 0.60 | 0.02 | 0.11 |

**Supplementary Table 3.** Relative energies ΔE [kJ (mol Si)⁻¹], density ρ [Si nm⁻³], average Si-O bond distances d(Si-O) [Å] and mean average deviations (MAD) of lattice parameters ΔL [Å] from experiments calculated at the SCAN level with two dispersion corrections (D3, MBD), and NNPscan level (SCAN+D3) for α-quartz (qu), α-cristobalite (cr), tridymite (tri) and purely siliceous zeolites. Experimental values for relative transition enthalpies ΔH and structural parameters were taken from Refs. 3–5.

|     | EXP |   |   | SCAN+D3 |   |   |   | SCAN+MBD |   |   |   | NNPscan |   |   |   |
| --- | --- | --- | --- | --- | --- | --- | --- | --- | --- | --- | --- | --- | --- | --- | --- |
|     | ΔH | ρ | d(Si-O) | ΔE | ρ | d(Si-O) | ΔL | ΔE | ρ | d(Si-O) | ΔL | ΔE | ρ | d(Si-O) | ΔL |
| qu  | 0.0 | 26.52 | 1.61 | 0.0 | 26.61 | 1.61 | 0.01 | 0.0 | 26.66 | 1.61 | 0.01 | 0.0 | 26.94 | 1.61 | 0.03 |
| cr  | 2.8 | 23.38 | 1.60 | 4.1 | 23.33 | 1.61 | 0.00 | 5.0 | 23.36 | 1.61 | 0.00 | 6.0 | 22.14 | 1.61 | 0.10 |
| tr  | 3.2 | 22.35 | 1.56 | 6.0 | 22.16 | 1.60 | 0.10 | 6.1 | 22.19 | 1.60 | 0.09 | 6.4 | 21.60 | 1.60 | 0.09 |
| AFI | 7.2 | 16.89 | 1.61 | 10.5 | 17.16 | 1.60 | 0.06 | 10.8 | 17.20 | 1.60 | 0.07 | 10.3 | 17.15 | 1.60 | 0.05 |
| FER | 6.6 | 17.55 | 1.61 | 9.8 | 17.90 | 1.60 | 0.09 | 9.6 | 17.94 | 1.60 | 0.10 | 10.8 | 17.85 | 1.60 | 0.08 |
| IFR | 10.0 | 17.15 | 1.61 | 11.0 | 17.16 | 1.61 | 0.03 | 11.6 | 17.13 | 1.61 | 0.03 | 11.3 | 16.86 | 1.61 | 0.06 |
| MTW | 8.7 | 18.23 | 1.61 | 9.4 | 18.50 | 1.60 | 0.06 | 8.9 | 18.47 | 1.60 | 0.05 | 10.0 | 18.28 | 1.60 | 0.02 |
| **MAD** | - | - | - | 2.1 | 0.18 | 0.01 | 0.05 | 2.3 | 0.18 | 0.01 | 0.05 | 2.7 | 0.47 | 0.01 | 0.06 |

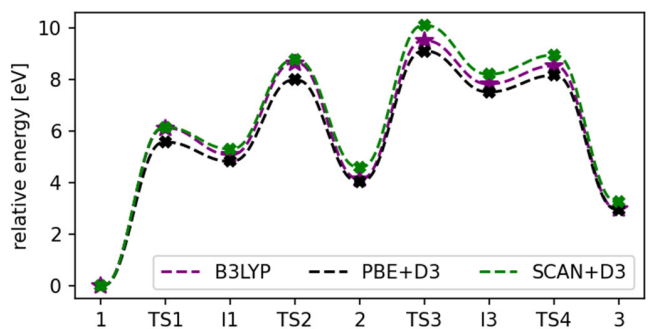

**Supplementary Figure 3** Comparison of B3LYP (taken from Ref. 6), SCAN+D3 and PBE+D3 energies for the Stone-Wales defect formation in a silica bilayer (see Fig. 5).

Supplementary Figure 4 depicts the energy-density plots $E(\rho)$ (*cf*. Fig. 3) of the IZA database calculated using SLC and NNPscan. It also shows the results of linear regression for 40 zeolites that exist in their purely siliceous form.[1] The obtained functions were shifted to indicate the accessible energy-density range of siliceous zeolite frameworks. NNPscan optimizations resulted in considerably lower relative energies than the SLC level calculations, particularly for three-ring containing zeolites such as Beryllosilicates (OBW, NAB, JOZ, OSO).[2] However, the Beryllosilicates and the 'unfeasible' high silica zeolite IPC-10[3] show relative energies above the accessible range of siliceous frameworks in both cases NNPscan and SLC.

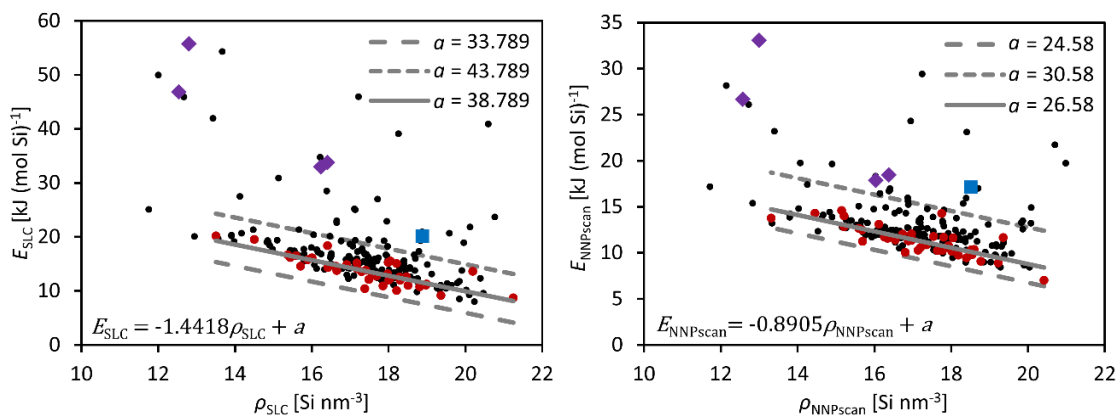

**Supplementary Figure 4** Relative energy $E$ as a function of framework density $\rho$ calculated at the SLC (left) and NNPscan (right) level of the IZA database (black dots). The solid lines show the linear regression of $E(\rho)$ for 40 purely siliceous zeolites (red dots). The obtained linear functions $E(\rho)$ were shifted by ±5 (SLC) and -2/+4 kJ mol$^{-1}$ (NNPscan) to define the accessible energy range of siliceous zeolites (dashed lines, *cf*. Fig. 3). Purple diamonds: Beryllosilicate zeolites (OBW, NAB, JOZ, OSO), blue square: IPC-10 (PCR).

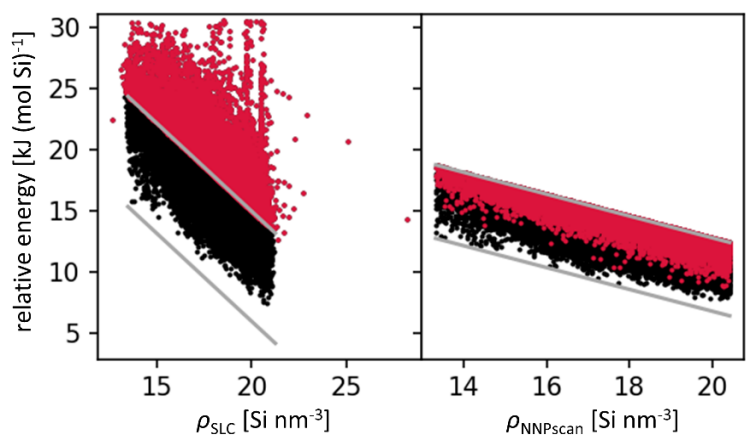

**Supplementary Figure 5** Relative energy $E$ as a function of framework density $\rho$ calculated at the SLC (left) and NNPscan (right) level. The solid lines indicate the thermodynamically accessible range of purely siliceous zeolites (*cf.* Figure 3 and S4). The black dots show the 32794 (SLC predicted) accessible frameworks. Red dots highlight the 20340 additional zeolites that are within the NNPscan predicted, thermodynamically accessible range of solvothermal synthesis routes.

**Supplementary Table 4.** Root mean square (RMSE) and mean absolute error (MAE) of energies [meV atom$^{-1}$] and forces [eV Å$^{-1}$] calculated for all test cases and only for equilibrium configurations (EQ) with respect to PBE+D3 results. NNP results include calculations of a single potential (sNNP) and an ensemble of six NNPs (eNNP).

| Level of theory | Energy (EQ) | | Forces (EQ) | | Energy (all) | | Forces (all) | |
|---|---|---|---|---|---|---|---|---|
| | MAE | RMSE | MAE | RMSE | MAE | RMSE | MAE | RMSE |
| PBE+D3 | 2.34 | 2.89 | 0.128 | 0.186 | 16.1 | 31.5 | 0.171 | 0.233 |
| sNNPpbe | 2.51 | 3.36 | 0.044 | 0.064 | 3.65 | 5.25 | 0.162 | 0.283 |
| eNNPpbe | 2.46 | 3.52 | 0.042 | 0.062 | 3.35 | 4.90 | 0.144 | 0.247 |

The test set used for error calculation in Table 1 did not include the structures from the second half of the ZA simulations since these are considered to be in an extrapolation region of the configuration space with energy errors between ca. 20-40 meV atom$^{-1}$ (*cf.* Fig. 6b and 6c).

**Supplementary Table 5.** Relative energies Δ$E$ [kJ (mol Si)$^{-1}$] and density $\rho$ [Si nm$^{-3}$] calculated at the DFT, NNP (single potential sNNP, ensemble average eNNP), and SLC level for α-quartz (qu), α-cristobalite (cr), tridymite (tri), moganite (mo), coesite (co), existing siliceous[a] and one hypothetical (HYP) zeolite. Experimental values for relative transition enthalpies Δ$H$ and densities were taken from Refs. 3–5. Mean absolute deviation (MAD) with respect to experimental values is given for all structures and a subset of structures optimized at the DFT level (MAD$_{sub}$).

|  | SLC | | SCAN+D3 | | sNNPscan | | eNNPscan | | PBE+D3 | | sNNPpbe | | eNNPpbe | | EXP | |
|---|---|---|---|---|---|---|---|---|---|---|---|---|---|---|---|---|
|  | Δ$E$ | $\rho$ | Δ$E$ | $\rho$ | Δ$E$ | $P$ | Δ$E$ | $\rho$ | Δ$E$ | $\rho$ | Δ$E$ | $\rho$ | Δ$E$ | $\rho$ | Δ$H$ | $\rho$ |
| qu | 0.00 | 27.74 | 0.00 | 26.61 | 0.00 | 26.99 | 0.00 | 26.87 | 0.00 | 25.92 | 0.00 | 25.11 | 0.00 | 25.72 | 0.00 | 26.52 |
| cr | 3.54 | 23.13 | 4.36 | 23.33 | 6.00 | 23.47 | 6.06 | 23.57 | 4.03 | 22.68 | 4.26 | 24.22 | 4.41 | 24.11 | 2.84 | 23.38 |
| tr | 5.41 | 21.91 | 7.08 | 22.16 | 6.42 | 21.75 | 6.44 | 21.62 | 6.49 | 21.61 | 7.07 | 21.75 | 6.70 | 21.42 | 3.21 | 22.35 |
| mo | 1.06 | 29.26 | 0.66 | 27.06 | -0.36 | 27.04 | -0.01 | 27.10 | 0.46 | 26.41 | 0.52 | 26.41 | -0.10 | 26.39 | 3.40 | 26.22 |
| co | 1.96 | 30.90 | 2.03 | 29.43 | 1.36 | 29.41 | 1.44 | 29.46 | 3.26 | 28.65 | 3.24 | 28.73 | 2.85 | 28.68 | 2.93 | 29.26 |
| AFI | 12.10 | 17.49 | 11.38 | 17.16 | 10.27 | 17.14 | 9.61 | 17.48 | 12.04 | 16.89 | 11.58 | 16.77 | 10.84 | 16.75 | 7.20 | 16.89 |
| FER | 11.94 | 18.30 | 10.64 | 17.90 | 10.77 | 17.94 | 10.17 | 18.26 | 11.20 | 17.46 | 11.33 | 18.22 | 11.29 | 17.93 | 6.60 | 17.55 |
| HYP | 5.61 | 26.57 | 3.64 | 26.35 | -0.46 | 26.34 | -0.23 | 26.26 | 2.90 | 25.70 | -0.59 | 25.50 | -0.75 | 25.49 | - | - |
| IFR | 15.17 | 17.17 | 12.04 | 17.16 | 11.31 | 16.88 | 11.47 | 16.91 | 12.03 | 17.15 | 12.01 | 16.78 | 11.51 | 16.74 | 10.00 | 17.15 |
| CFI | 13.56 | 17.30 | 12.22 | 16.98 | 12.08 | 16.97 | 10.41 | 17.69 | 13.46 | 16.77 | 14.23 | 16.54 | 13.54 | 16.51 | 8.80 | 16.77 |
| MTW | 11.02 | 18.51 | 9.89 | 18.49 | 9.97 | 18.54 | 9.58 | 18.49 | 10.89 | 18.23 | 11.40 | 18.17 | 10.25 | 18.35 | 8.70 | 18.23 |
| MEL | 10.92 | 17.85 | - | - | 10.23 | 17.64 | 9.86 | 17.66 | - | - | 11.00 | 17.67 | 10.40 | 17.63 | 8.20 | 17.80 |
| MWW | 14.56 | 16.62 | - | - | 11.38 | 16.54 | 11.54 | 16.52 | - | - | 12.45 | 16.35 | 12.03 | 16.30 | 10.40 | 16.51 |
| ITE | 14.32 | 16.42 | - | - | 11.72 | 16.26 | 11.72 | 16.22 | - | - | 12.81 | 15.98 | 12.37 | 15.96 | 10.10 | 16.26 |
| AST | 18.35 | 16.41 | - | - | 13.09 | 16.12 | 12.29 | 16.93 | - | - | 15.22 | 15.76 | 12.84 | 16.96 | 10.90 | 17.29 |
| STT | 14.89 | 16.92 | - | - | 12.03 | 16.73 | 12.10 | 16.68 | - | - | 13.33 | 16.47 | 12.78 | 16.45 | 9.20 | 16.83 |
| CHA | 16.52 | 15.59 | - | - | 12.83 | 15.21 | 12.80 | 15.22 | - | - | 14.03 | 14.85 | 13.57 | 14.87 | 11.40 | 15.40 |
| BEA | 14.59 | 15.71 | - | - | 12.65 | 15.60 | 12.35 | 15.53 | - | - | 14.03 | 15.39 | 13.33 | 15.31 | 9.30 | 15.60 |
| MFI | 10.06 | 18.21 | - | - | 9.73 | 18.19 | 9.68 | 18.01 | - | - | 10.82 | 18.18 | 10.28 | 18.05 | 6.80 | 17.97 |
| ISV | 16.65 | 15.38 | - | - | 14.60 | 15.15 | 14.49 | 15.13 | - | - | 15.76 | 15.17 | 15.14 | 15.07 | 14.40 | 15.36 |
| FAU | 20.16 | 13.51 | - | - | 13.78 | 13.31 | 13.92 | 13.29 | - | - | 14.54 | 13.00 | 13.98 | 13.05 | 13.60 | 13.30 |
| **MADsub** | **3.19** | **0.90** | **2.66** | **0.25** | **2.76** | **0.36** | **2.37** | **0.51** | **2.90** | **0.29** | **3.08** | **0.50** | **2.75** | **0.45** | - | - |
| **MAD** | **3.97** | **0.55** | - | - | **2.24** | **0.35** | **1.99** | **0.28** | - | - | **3.02** | **0.51** | **2.50** | **0.46** | - | - |

[a] AFI: Aluminophosphate-five; FER: Ferrierite; IFR: Instituto de Tecnologia Quimica Valencia - four; CFI: California Institute of Technology - five; MTW: Zeolite Socony Mobil - twelve; MEL: Zeolite Socony Mobil - eleven; MWW: Mobil Composition of Matter-twenty-two; ITE: Instituto de Tecnologia Quimica Valencia - three; AST: AlPO$_4$-16 (sixteen); STT: Standard Oil Synthetic Zeolite - twenty-three; CHA: Chabazite; BEA: Zeolite Beta; MFI: Zeolite Socony Mobil - five; ISV: Instituto de Tecnologia Quimica Valencia - seven; FAU: Faujasite

**Supplementary Table 6.** Frequencies [THz] and irreproducible representations (irrep) of the vibrational modes of α-cristobalite obtained from experiment (exp)[7–10] as well as simulations (SCAN+D3, NNPscan) using finite differences (FD) and MD trajectories. MADs are given with respect to experiments.

| irrep | exp | irrep | SCAN+D3 (harm) | NNPscan (harm) | NNPscan (MD) |
|---|---|---|---|---|---|
| E | 35.9[c] | E | 36.0 | 34.3 | 35.4 |
| B$_2$ | 35.6[b] | B$_2$ | 34.0 | 33.2 | 34.3 |
| A$_2$ | 34.3[a] | A$_2$ | 32.9 | 31.3 | 32.3 |
| E | 33.0[a] | E | 32.7 | 31.3 | 32.4 |
| A$_1$ or B$_1$ | 32.6[b] | B$_1$ | 32.7 | 31.3 | 32.4 |
|  | 32.3[b] | A$_1$ | 32.6 | 31.0 | 32.2 |
| n/a | 23.9[a] | A$_2$ | 23.8 | 23.5 | 24.0 |
|  | 23.9[b] | B$_2$ | 23.8 | 23.4 | 24.0 |
| B$_1$ | 23.5[b] | B$_1$ | 23.7 | 22.8 | 23.1 |
| E | 18.7[a] | E | 18.9 | 17.5 | 18.3 |
| A$_2$ | 14.8[a] | A$_2$ | 15.2 | 13.8 | 14.1 |
| E | 14.4[a] | E | 14.5 | 13.2 | 13.7 |
| A$_1$ or B$_2$ | 12.8[b] | A$_1$ | 13.2 | 12.6 | 13.1 |
|  |  | B$_2$ | 13.1 | 13.6 | 13.8 |
| E | 11.4[c] | E | 11.8 | 11.1 | 11.5 |
| A$_1$ or B$_1$ | 11.0[b] | A$_1$ | 11.2 | 11.9 | 11.1 |
|  |  | B$_1$ | 11.6 | 9.9 | 10.4 |
| A$_2$ | 9.0[a] | A$_2$ | 8.9 | 8.8 | 9.1 |
| B$_2$ | 8.6[b] | B$_2$ | 8.6 | 8.4 | 8.8 |
| E | 8.3[c] | E | 8.3 | 8.2 | 8.6 |
| A$_1$ | 7.0[b] | A$_1$ | 7.2 | 7.5 | 7.2 |
| E | 4.4[a] | E | 5.0 | 4.5 | 4.2 |
| B$_1$ | 3.6[b] | B$_1$ | 3.8 | 3.9 | 3.2 |
| B$_1$ | 1.5[b] | B$_1$ | 1.9 | 1.1 | 0.8 |
| **MAD** | - | - | **0.3** | **0.9** | **0.5** |

Supplementary Figure 6 shows the vibrational density of states (VDOS) of silica glass models obtained from three independent simulated annealing runs. The excellent agreement of all VDOS indicates that the simulation procedure yielded reliable and reproducible glass structure predictions.

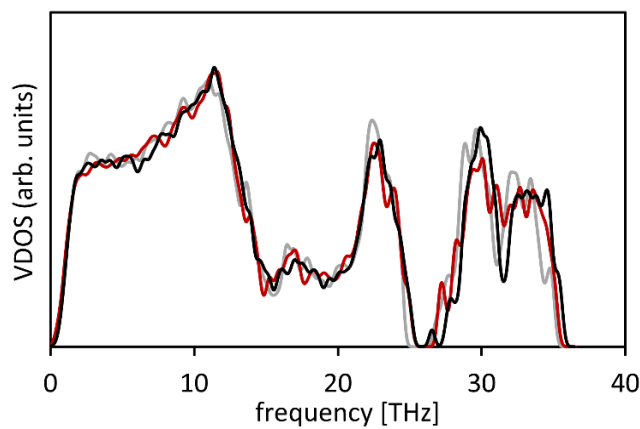

**Supplementary Figure 6** Vibrational density of states (VDOS) of amorphous silica structures obtained from three independent simulated annealing runs at the NNPscan level.

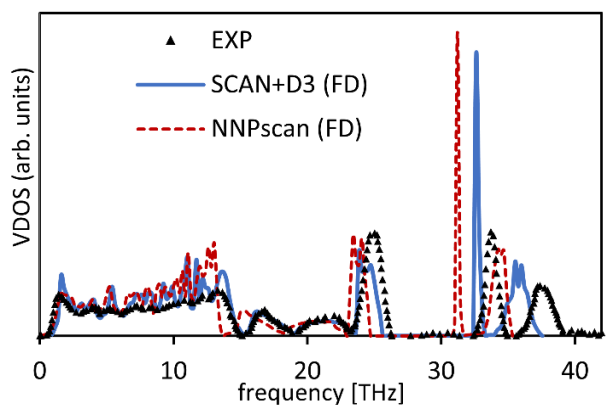

**Supplementary Figure 7** Vibrational density of states (VDOS) of α-cristobalite obtained from experiments (EXP), SCAN+D3 and NNPscan calculations using the finite-difference (FD) approach (*cf.* Figure 4).

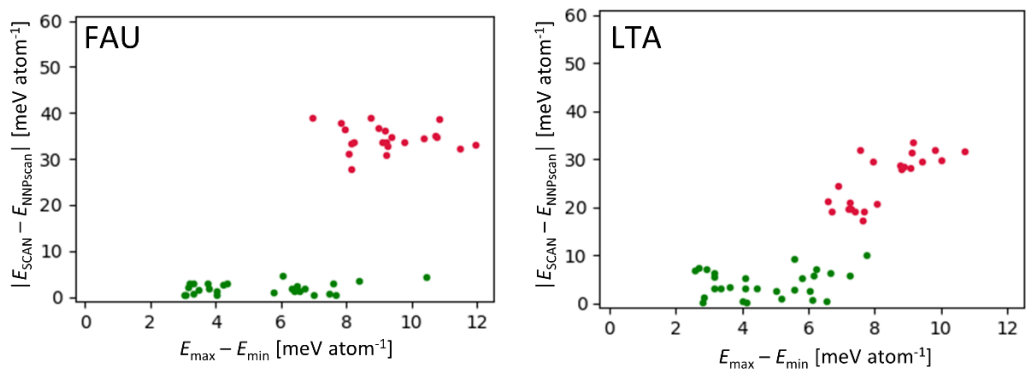

**Supplementary Figure 8** NNP errors with the energy prediction spread (difference of max. and min. energy) of the NNP ensemble obtained from zeolite amorphization simulations (*cf.* Figure 6). Structures in the interpolated region are in green and extrapolated ones in red.

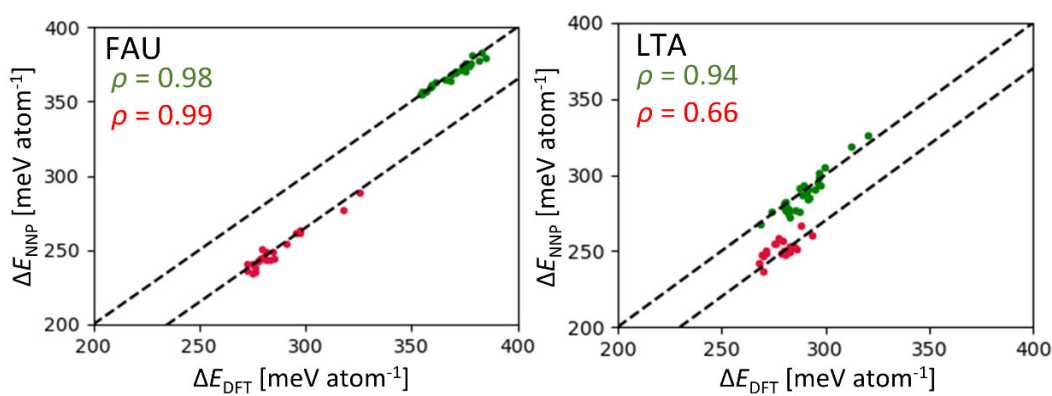

**Supplementary Figure 9** Qualitative correlation of relative NNP and DFT energies (green: interpolated, red: extrapolated energies) obtained from zeolite amorphization simulations (*cf.* Figure 6) along with Pearson correlation coefficients $\rho$.

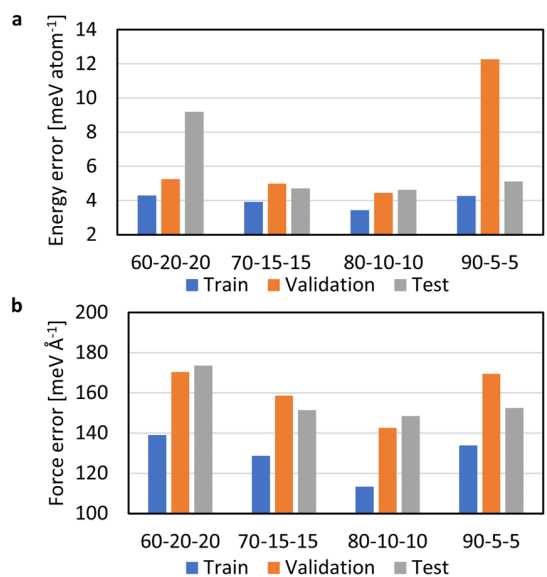

**Supplementary Figure 10 a** NNP energy and **b** force errors (RMSE) of the training, validation, and test part of the SCAN+D3 reference database as a function of the train-validation-test splits [%].

## 3. Supplementary references